\documentclass[aip,pop,preprint,superscriptaddress,amsmath,amsfonts,amssymb]{revtex4-1}

\usepackage{amsfonts}
\usepackage{amsmath}
\usepackage{bm}

\usepackage[pdftex]{color,graphicx}
\usepackage[colorlinks,linkcolor=blue,citecolor=blue,bookmarks,pdfstartview=FitH]{hyperref}
\usepackage{graphics}

\newcommand{\gav}[1]{\left \langle #1\right \rangle}
\newcommand{\sav}[1]{\left \langle #1\right \rangle_{\psi}}
\newcommand{\vav}[1]{\left \langle #1\right \rangle_{v}}
\newcommand{\wb}[1]{\bm #1}

	\newcommand{\cb}{\color{blue} }
\begin{document}

\title{Gyrokinetic theory for particle transport in fusion plasmas}
\author{Matteo Valerio Falessi}
\email{matteo.falessi@enea.it}
\affiliation{ENEA, Fusion and Nuclear Safety Department, C. R. Frascati, Via E. Fermi 45, 00044 Frascati (Roma), Italy}
\author{Fulvio Zonca}
\affiliation{ENEA, Fusion and Nuclear Safety Department, C. R. Frascati, Via E. Fermi 45, 00044 Frascati (Roma), Italy}
\affiliation{Institute for Fusion Theory and Simulation and Department of Physics,
Zhejiang University, Hangzhou 310027, China}
\begin{abstract}
A set of equations is derived describing the macroscopic transport of particles and energy in a thermonuclear plasma on the energy confinement time.
The equations thus derived allow studying collisional and turbulent transport self-consistently, retaining the effect of magnetic field geometry without postulating any scale separation between the reference state and fluctuations.
Previously, assuming scale separation, transport equations have been derived from kinetic equations by means of multiple-scale perturbation analysis and spatio-temporal averaging.
In this work, the evolution equations for the moments of the distribution function are obtained following the standard approach; meanwhile, gyrokinetic theory has been used to explicitly express the fluctuation induced fluxes.
In this way, equations for the transport of particles and energy up to the transport time scale can be derived using standard first order gyrokinetics.
\end{abstract}
\maketitle

\section{Introduction}
 \label{sec:intr}
 Predicting the dynamics of a thermonuclear plasma is fundamental in order to make nuclear fusion a reliable source of energy. A precise and quantitative definition of what this means essentially implies discussing and understanding the spatiotemporal scales of nonlinear evolution of plasma profiles due to Coulomb collisions and fluctuating fields on the same footing. This also implies a proper definition of a plasma reference state and of the deviation of the system from that "average" condition.

In collisionless fusion plasmas the turbulent fluctuation spectrum is dominated by frequencies usually larger than the inverse characteristic collisional relaxation time. Meanwhile, due to the assumption of strong magnetization, transport parallel to the ambient magnetic field is much faster than across magnetic flux surfaces. Therefore, transport theory in fusion plasmas has been developed as intrinsically local in order to describe the evolution of the reference state on a short timescale, dominated by fluctuation induced cross-field transport (across equilibrium magnetic flux surfaces) on the micro-scales characterizing drift wave turbulence. The study of plasma global evolution has been often pursued by means of analysis based on ad hoc models. Only a limited number of works have tackled self-consistently collisional and fluctuation induced transport, e.g. Refs. \citenum{abel2013multiscale,plunk2009theory,sugama1996transport,balescu1990anomalous,shaing1988neoclassical}. A practical and effective approach to this problem intuitively yields to defining suitable time and spatial averages. In fact, plasma reference state should evolve slowly in some sense, while cross field transport should cause the distortion of plasma profiles on a sufficiently long length scale. The derivation of a set of equations describing the plasma global evolution on a given time scale is the main requirement to reach this goal. The self-consistency of the adopted description is of fundamental importance in order to understand transport processes on a given time scale because of the possible interplay between collisions and turbulence. In turn, this is crucial in order to predict fluxes of particle and energy and, ultimately, the overall plasma evolution.

The aim of this work is the self-consistent study of transport processes in a thermonuclear plasma on the energy confinement time scale. We adopt a moment equation approach to transport equations \cite{hinton1976theory}, combined with the conceptual framework of nonlinear gyrokinetic theory \cite{frieman1982nonlinear,brizard2007foundations}. In this way, we provide a compact and physically transparent derivation of cross-field particle and energy fluxes, which include collisional and fluctuation induced transport processes on the same footing. The resulting transport equations generalize and extend the analysis of Ref. \citenum{hinton1976theory} to fluctuation induced transport. More precisely, fluctuations generally describe the deviation of the plasma from the reference state, which evolves in time consistently with the fluctuation spectrum and is characterized by spatiotemporal scales that are consistent with the so-called transport ordering typical of strongly magnetized fusion plasmas \cite{hinton1976theory}. Meanwhile, the fluctuation spectrum is assumed to be generic and consistently imposed with the gyrokinetic ordering of spatiotemporal scales. Fluctuation induced fluxes are described using gyrokinetic field theory \cite{brizard2007foundations} and are expressed in terms of the gyrocenter distribution function. The main advantage of studying the evolution equations for the moments instead of directly writing the kinetic expression of fluxes in terms of the particle distribution function is that we derive the equations for the fluctuations induced fluxes up to the transport time scale using standard first order gyrokinetic theory. Our analysis also recovers, in the appropriate limit, the results originally proposed in Refs. \citenum{plunk2009theory,barnes2009trinity,abel2013multiscale}. 

The work is organized as follows. In Section \ref{sec:fund}, we provide a description of the notation and of the ordering of the physical quantities. In Sections \ref{sec:particletransp} and \ref{sec:energytransp}, we extend the analysis of Ref. \citenum{hinton1976theory} to fluctuation induced transport obtaining expressions for collisional and fluctuation induced fluxes that are valid up to the energy confinement time scale. Section \ref{sec:fluctindfluxes} contains the derivation of the fluctuation induced fluxes in terms of the gyrocenter distribution function. In section \ref{sec:longer}, we briefly discuss the further analysis required to derive a set of transport equations valid on longer timescales. Final conclusions and discussions are given in Section \ref{sec:concl}.

\section{Theoretical framework and ordering assumptions}
\label{sec:fund}
In this work, we study transport processes in the core of a thermonuclear plasma and, therefore, we can assume that the plasma is strongly magnetized. Thus, the particle distribution function can be written as the sum of a reference distribution $F_{0}$ function and a small perturbation $\delta f$:
   \begin{equation}
   \label{eq:6}
   f=F_{0}+\delta f {\cb{\; ,}}
   \end{equation}
   where the characteristic length scale of variation of the reference distribution function, i.e. $F_{0}$, is such that $L \sim \delta^{-1}\rho$ where $\rho$ is the Larmor radius. Following Refs. \citenum{hinton1976theory,frieman1982nonlinear}, we  assume that reference states evolve sufficiently slowly and are characterized by macroscopic profiles (conventionally denoted as $p_0$) satisfying the following scaling assumption:
   \begin{equation}
   \label{eq:1}
   \omega^{-1}\frac{\partial }{\partial t}\ln{p_{0}}= \omega^{-1}p_{0}^{-1}\frac{\partial p_{0}}{\partial t}  \sim {\cal O}(\delta^{2}) {\cb{ \; ,}}
   \end{equation}
   where $p_{0}$ stands for $n_{0}$ {\em etc.}, and we use the notation for the moments of the distribution function introduced in Ref.\citenum{hazeltine2003plasma}. We further assume the so-called drift ordering:
   \begin{equation}
   \label{eq:5}
   \frac{c E}{B_{0} v_{th}}\sim {\cal O}(\delta) {\cb{\; ,}}
   \end{equation}
 where $v_{th}$ is the particle thermal speed, and other symbols are standard.
   Consistently with this standard approach \cite{hinton1976theory}, we write electromagnetic fields as the sum of reference fields, self-consistently determined within the reference state, varying on the equilibrium lengthscale $L$, and fluctuations. Following Ref.\citenum{frieman1982nonlinear}, we adopt the gyrokinetic ordering for fluctuating quantities:
   \begin{equation}
     \label{eq:n2}
     \frac{|\partial/\partial t|}{|\Omega|}\sim \left | \frac{\delta B}{B_{0}}\right |\sim\frac{{\wb \nabla}_{\parallel}}{{\wb \nabla}_{\perp}} \sim \frac{{\wb k_{\parallel}}}{{\wb k}_{\perp}} \sim \mathcal{O}(\delta) {\cb{\; ,}}
   \end{equation}
where $\Omega$ is the particle cyclotron frequency in the reference state magnetic field.
We also adopt straight magnetic field line toroidal flux coordinates, see e.g. Ref. \citenum{d2012flux}, and we assume axisymmetry of the reference state. Therefore, without loss of generality, the reference magnetic field  has the following expression:
   \begin{equation}
\label{eq:37}
{\bm B}_{0}= F {\bm \nabla}\phi + {\bm \nabla}\phi \times {\bm\nabla}\psi.
\end{equation}
With these assumptions, the distribution function can be assumed as Maxwellian at the leading order:
   \begin{equation}
   \label{eq:6a}
   f = f_{M} + {\cal O}(\delta),  \quad f_{M}= n_{0}(\pi^{1/2}v_{th })^{-3}e^{-(v/v_{th})^{2}} {\cb{\; .}}
  \end{equation}
Following Ref. \citenum{hinton1976theory}, we assume that the parallel flow is strongly subsonic and that there is small pressure anisotropy between the directions perpendicular and parallel to ${\wb B}_{0}$ due to Coulomb collisions. Thus, we obtain:
  \begin{equation}
  \label{eq:7}
  \left\{ n \bm V, \bm F, \bm Q, [ \bm P - \bm I p ], [ \bm R - \bm I (5/2) p (T/m) ]\right\} \sim {\cal O}(\delta).
  \end{equation}
  Here, again, the notation for moments of the particle distribution function is that of Ref. \citenum{hazeltine2003plasma};
  {\em i.e.},  $n \bm V$ is the particle flux, $\bm F$ is the friction force, $\bm Q$ is the energy flux, $\bm P$ is the stress tensor,
  and $\bm R$ is the energy-weighted stress tensor. Meanwhile, with standard notation, $\bm I$ denotes the 
  unit diagonal matrix. Furthermore, space and time scales are normalized to $|\rho|$ and $|\Omega^{-1}|$, respectively, density is expressed in units of its local equilibrium value, etc. Consistent with this, we obtain{\cb{\cite{hinton1976theory}}}:
  \begin{equation}
  \label{eq:9}
  \bm E=- \bm \nabla \Phi + \cal{O}(\delta)
  \end{equation}
  and, thus, a consequence of this ordering is that the reference state electric field is mainly electrostatic and 
  satisfies the drift ordering of Eq. (\ref{eq:5}). From the lowest order parallel force balance equation and  parallel energy flux conservation,  it follows that $\bm b \cdot \bm \nabla n = {\cal O}(\delta )$ and $\bm b \cdot \bm \nabla \Phi = {\cal O}(\delta)$ with $\bm b = \bm B_{0} /B_{0} $. That is, at the lowest order, temperature and density are constant along magnetic field lines of the reference state and, because of their ergodic properties \cite{hinton1976theory}, they are also constant on magnetic surfaces.

\section{Particle transport}
\label{sec:particletransp}

In this section, as anticipated in Secs. \ref{sec:intr} and \ref{sec:fund}, we adopt the same approach and theoretical framework of Ref. \citenum{hinton1976theory},
with the additional element of taking into account the effect of fluctuating fields on transport processes. The fluctuation spectrum is generic, but assumed to be
consistent with the gyrokinetic ordering of spatiotemporal scales given by Eq. (\ref{eq:n2}).

In the following, we proceed in the derivation of the equation describing particle transport up to the energy confinement time by acting with the projection operator $R^2 \bm \nabla \phi \, \cdot$ on the momentum equation \cite{hinton1976theory}. Adopting the theoretical framework and ordering introduced in  {\cb{Sec. \ref{sec:fund}}}, it can be shown that the ${\bm P}$ tensor is symmetric up to $\mathcal{O}(\delta)$ and, thus, due to the anti-symmetry of $\bm \nabla ( R^2 \bm \nabla \phi )$, we obtain the following expression:
\begin{eqnarray}
\label{eq:93}
\left\langle R^2 \bm \nabla \phi \cdot \partial_t (n m \bm V ) \right\rangle_\psi & + & \frac{1}{V'} \frac{\partial}{\partial \psi} \left\langle V' \bm \nabla \psi \cdot \bm P \cdot R^2 \bm \nabla \phi\right\rangle_\psi  = \left\langle R^2 \bm \nabla \phi \cdot(en \bm E + \bm F ) \right\rangle_\psi+ \\
\nonumber & & + \left\langle R^2 \bm \nabla \phi \cdot \left(e n /c {\bm V} \times {\bm B} \right) \right\rangle_\psi .
\end{eqnarray}
Note that, here, the electric field $\bm E = \bm E_0 + \delta \bm E$ satisfies the drift ordering of Eq. (\ref{eq:5}). 
We now use the identity:
\begin{equation}
\label{eq:94}
\bm B_0 \times R^2 \bm \nabla \phi = \bm \nabla \psi,
\end{equation}
in order to obtain the following relation, which is valid up to ${\cal O}(\delta \omega n_{0}m v_{th}L)$:
\begin{eqnarray}
\label{eq:96}
\left\langle (en/c) \bm V \cdot \nabla \psi \right\rangle_\psi  & = & - \left\langle (en \bm E + \bm F ) \cdot R^2 \bm \nabla \phi \right\rangle_\psi +
\\\nonumber  & & + \left\langle (en/c) \bm V \cdot (\bm B_0 - \bm B) \times R^2 \bm \nabla \phi \right\rangle_\psi.
\end{eqnarray}
This is the analogue of Eq. (2.93) in Ref. \citenum{hinton1976theory}, where we have considered also the contribution of the fluctuating fields. We note that the term $\left\langle R^2 \bm \nabla \phi \cdot \partial_t (n m \bm V ) \right\rangle_\psi$ is generally non-negligible in the presence of fluctuations. In particular, since $\sav{\partial_t \left(n \bm V\right)} \sim \bm V \sav{\partial_t n}$, we can estimate its magnitude from the surface averaged continuity equation. Nonlinear (fluctuation-induced) terms may generally have characteristic length-scale in between those of the turbulent fluctuation spectrum and of reference profiles; that is, mesoscale structures. Thus, an approach that postulates a systematic scale separation between fluctuation-induced profile distortions and reference profiles is questionable. Labeling the characteristic length-scale of these  structures as $k_{z}^{-1}$, we obtain $\sav{\partial_{t} n} \sim \delta^{2}k_{z}n_{0}v_{th}$. On the characteristic lengthscale of reference profiles, the evolution of the density  is obtained by letting $k_{z}^{-1}\sim L$  and, therefore,  Eq. (\ref{eq:96}) follows. From this argument, we deduce that studying the evolution equations for the moments of the distribution function,  a fluid approach instead of a kinetic description, is particularly convenient only on the length-scale of the reference profiles, while it gets increasingly more difficult, although in principle feasible, on smaller length{\cb{-}}scales. For this reason, in the present work we investigate how the evolution 
of reference states on the macroscales is affected by a prescribed generic spectrum of (gyrokinetic) fluctuations. In other words, fluctuations themselves can be generally microscopic, consistent with Eq. (\ref{eq:n2}), but their effect on profile evolution are computed on the characteristic spatiotemporal scales 
of the reference states themselves; that is, the possible formation of mesoscale structure will be ignored. We will study transport on an arbitrary length-scales (including mesoscales) in a future work, using directly the nonlinear gyrokinetic equation (see Ref. \citenum{falessi2017gyrokinetic}).
The expression for the particle flux, Eq.(\ref{eq:96}), can be used to compute the evolution of the density profile in the continuity equation. Using this method, we can describe the fluxes up to second order using the information on the distribution function accurate up to first order \cite{hinton1976theory}. Equation (\ref{eq:96}) includes classical, neoclassical and fluctuation-induced transport and, therefore, generalizes the result derived in Ref. \citenum{hinton1976theory}. Using the following relation:
\begin{equation}
\label{eq:99}
{\bm b} \times {\bm\nabla} \psi = F {\bm b} - B R^2 {\bm \nabla}\phi,
\end{equation}
starting from the expression for the fluxes derived in Ref. \citenum{hinton1976theory}, we can identify the classical (subscript ``$c$'') and neoclassical 
(subscript ``$NC$'') contributions in Eq. (\ref{eq:96}), obtaining:
\begin{equation}
\label{eq:97}
 \left\langle (en/c) \bm V \cdot \nabla \psi \right\rangle_{\psi c} = - \left\langle \bm F_\perp \cdot R^2 \bm \nabla \phi \right\rangle_\psi, \, \, \left\langle (en/c) \bm V \cdot \nabla \psi \right\rangle_{\psi NC} = - \left\langle ( en \bm E_0 + \bm F_\parallel ) \cdot R^2 \bm \nabla \phi \right\rangle_\psi.
\end{equation}
The distinction of classical and neoclassical fluxes is somewhat conventional, \emph{e.g.}, see Refs. \citenum{hinton1976theory,wimmel1970energy}, since it ultimately resorts to the effect of Coulomb collisions. The remaining terms of Eq. (\ref{eq:96}), which can be attributed to fluctuations (subscript ``$gk$'') as they vanish in their absence, read:
\begin{eqnarray}
\label{eq:101}
\left\langle (en/c) \bm V \cdot \nabla \psi \right\rangle_{\psi gk} & = & - \left\langle (en \bm E - en \bm E_0 ) \cdot R^2 \bm \nabla \phi \right\rangle_\psi
\nonumber \\ & & + \left\langle (en /c) \bm V \cdot (\bm B_0 - \bm B) \times R^2 \bm \nabla \phi \right\rangle_\psi {\cb{\; .}} 
\end{eqnarray}
Collecting the various contributions derived above, the density transport
equation can be written as:
\begin{equation}
\label{eq:116}
\left\langle\left\langle \partial_t f\right\rangle_v\right\rangle_\psi = - \frac{1}{V'} \frac{\partial}{\partial \psi} \left[ V' \left\langle n \bm V \cdot \nabla \psi \right\rangle_{\psi c} + V' \left\langle n \bm V \cdot \nabla \psi \right\rangle_{\psi NC} + V' \left\langle n \bm V \cdot \nabla \psi \right\rangle_{\psi gk} \right]
\end{equation}
where $V'=d V/d\psi$. This additive form does not imply that transport processes are independent
of each other and it is a mere consequence of the formal classification adopted here. It is readily recognized that the neoclassical flux in Eq. (\ref{eq:116}) could also depend on fluctuations intensity, although at higher (negligible) order, (see, {\em e.g.}, Ref.\citenum{sugama1996entropy}). Exploring transport processes more in depth, fluctuations may enhance the deviation of system from local thermodynamic equilibrium and cause structure formation in the particle phase space,  \emph{e.g.}, see Refs. \citenum{chen2007theory,zonca2006physics,zonca2015nonlinear}, which are eventually damped by collisions (enhanced collisional damping). Meanwhile, collisions may damp long lived structures formed by saturated instabilities, such as zonal flows   \cite{hasegawa1979nonlinear,lin1998turbulent,rosenbluth1998poloidal,hinton1999dynamics,diamond2005zonal,itoh2006physics}, or more generally zonal structures  \cite{zonca2006physics,chen01,guzdar01,gruzinov02,chen12}, which, in turn, regulate turbulent transport itself. We conclude this section by stressing that Eq. (\ref{eq:116}) describes all radial particle transport processes on the  characteristic length-scale of the reference state, and it is predictive on the energy confinement time-scale.

\section{Energy transport}
\label{sec:energytransp}
  Using the same  theoretical framework and approach of Sec. \ref{sec:particletransp}, we can derive an expression for the radial energy transport on the characteristic spatiotemporal scales of the reference states themselves. Taking the dot product of $R^{2}\wb{\nabla}\phi$ with the energy transport equation and taking the flux surface average yields \cite{hinton1976theory}:
  \begin{align*}
  \label{eq:108}
    \left \langle \frac{\partial {\bm Q}}{\partial t}\cdot R^2 \wb{\nabla}\phi \right \rangle_\psi + \left \langle{\bm \nabla} \cdot {\bm R}\cdot R^2 \wb{\nabla} \phi\right \rangle_\psi &- \left \langle \frac{e}{m}{\bm E}\cdot \left(\right. {\bm P} - p_\perp \bm I \left.\right) \cdot R^2{\bm \nabla}\phi \right \rangle_\psi +\\ \nonumber - \left \langle
    \frac{e}{m}{\bm E} \cdot
    {\bm I}\left( 2 p_\perp + \frac{1}{2}p_\parallel + \frac{m}{2} n V^2\right) \cdot R^2 {\bm \nabla}\phi \right \rangle_\psi 
    &- \left \langle \frac{e}{mc} {\bm Q} \times {\bm B} \cdot R^2\cdot {\bm \nabla}\phi\right \rangle = \left \langle {\bm G} \cdot R^2 {\bm \nabla}\phi\right \rangle.
  \end{align*}
Here, for the sake of generality, we have assumed that the leading order stress tensor is given in the Chew-Golberger-Low form \cite{chew56},
  $\bm P_{CGL} = p_\perp \bm I + (p_\parallel - p_\perp) \bm b \bm b$, having in mind the application to collisionless fusion plasmas; and $\bm G$ denotes
  the collisional change in the energy flux\cite{hinton1976theory}. Proceeding as in Sec. \ref{sec:particletransp} and applying the drift ordering, we obtain, up to order $\mathcal{O}(\delta^{2})$:
  \begin{equation}
  \label{eq:129}
  \left\langle \bm Q \cdot \nabla \psi \right\rangle_\psi = \left\langle \bm Q \cdot \nabla \psi \right\rangle_{\psi c} + \left\langle \bm Q \cdot \nabla \psi \right\rangle_{\psi NC} + \left\langle \bm Q \cdot \nabla \psi \right\rangle_{\psi gk} {\cb{\; ,}}
  \end{equation}
  where:
  \begin{align}
  \label{eq:144}
    \left\langle \bm Q \cdot \nabla \psi \right\rangle_{\psi c} &=  - (mc/e)\left\langle \bm G_\perp \cdot R^2 \bm \nabla \phi \right\rangle_\psi \\
    \left\langle \bm Q \cdot \nabla \psi \right\rangle_{\psi NC} &=  - \left\langle (c \bm E_0 (2 p_\perp + p_\parallel/2) + (mc/e) \bm G_\parallel ) \cdot R^2 \bm \nabla \phi \right\rangle_\psi\\
    \left\langle \bm Q \cdot \nabla \psi \right\rangle_{\psi gk}  = & - \left\langle c (\bm E - \bm E_0 ) (2 p_\perp + p_\parallel/2)  \cdot R^2 \bm \nabla \phi \right\rangle_\psi
    - \left\langle  \bm Q \cdot (\bm B - \bm B_0) \times R^2 \bm \nabla \phi \right\rangle_\psi.
  \end{align}
  The fluxes expressions are similar to the particle fluxes derived in the previous section. In order to compare these results
, we note that the energy transport equation can be cast in the following form:
  \begin{equation}
  \label{eq:149}
  \partial_t \left( p_\perp  + p_\parallel/2  \right)  + \bm \nabla \cdot \bm Q   = W + \left( e n\bm E + \bm F \right) \cdot \bm V - \partial_t \left( \frac{m}{2} n V^2 \right)
  {\cb{\; ,}}
  \end{equation}
  with $W$ representing the collisional energy exchange (denoted as $Q$ in Ref. \citenum{hinton1976theory}). Equation (\ref{eq:149}) can be re-written at the required order by using the leading order expression for ${\bm V}$:
  \begin{equation}
  \label{eq:154}
  \partial_t \left\langle p_\perp  + p_\parallel/2  \right\rangle_\psi  + \frac{1}{V'}
   \frac{\partial}{\partial \psi} \left[ V' \left(  \left\langle \bm Q \cdot  \bm \nabla \psi \right\rangle_\psi  + \left\langle c p_\perp \bm E \cdot R^2 \bm \nabla \phi \right\rangle_\psi  \right) \right]   = \left\langle W \right\rangle_\psi
   \; ;
  \end{equation}
  and, thus, we have demonstrated that, at the relevant order in our asymptotic expansion in the drift parameter, the evolution equation for $(p_{\perp}+p_{\parallel}/2)$ is a transport equation with a collisional heating source $\left\langle W \right\rangle_\psi$ and with an effective radial flux:
  \begin{equation}
  \label{eq:155}
  \left\langle \bm Q_{\mathrm eff} \cdot  \bm \nabla \psi \right\rangle_\psi \equiv \left\langle \bm Q \cdot  \bm \nabla \psi \right\rangle_\psi  + \left\langle c p_\perp \bm E \cdot R^2 \bm \nabla \phi \right\rangle_\psi
  \; .
  \end{equation}
Using this result we can write the expressions for the effective fluxes to be used in the energy evolution equation:
  \begin{align}
  \label{eq:157}
  \left\langle \bm Q_{\mathrm eff} \cdot \nabla \psi \right\rangle_{\psi c} & = - (mc/e)\left\langle \bm G_\perp \cdot R^2 \bm \nabla \phi \right\rangle_\psi \\  \left\langle \bm Q_{\mathrm eff} \cdot \nabla \psi \right\rangle_{\psi NC} & = - \left\langle (c \bm E_0 (p_\perp + p_\parallel/2) + (mc/e) \bm G_\parallel ) \cdot R^2 \bm \nabla \phi \right\rangle_\psi \\  \left\langle \bm Q_{\mathrm eff} \cdot \nabla \psi \right\rangle_{\psi gk} & = - \left\langle c (\bm E - \bm E_0 ) (p_\perp + p_\parallel/2)  \cdot R^2 \bm \nabla \phi \right\rangle_\psi
    - \left\langle  \bm Q \cdot (\bm B - \bm B_0) \times R^2 \bm \nabla \phi \right\rangle_\psi.
  \end{align}

By direct comparison with the collisional and gyrokinetic particle fluxes, we readily see that the expression are formally the same, with energy fluxes weighted by $m v^{2}/2$. This is consistent with the results obtained directly from kinetic theory, see Ref. \citenum{falessi2017gyrokinetic}, which will be analyzed in a future work.

\section{Fluctuation induced fluxes}
\label{sec:fluctindfluxes}
The explicit calculation of the fluxes derived in the previous sections requires neoclassical transport theory, \emph{e.g}., see Ref. \citenum{hinton1976theory}, and nonlinear gyrokinetic theory, \emph{e.g.}, see Ref. \citenum{brizard2007foundations,frieman1982nonlinear}. In this work, we calculate the explicit expressions of fluctuation induced fluxes in terms of the first order distribution function, adopting the 
well-known perturbation expansion assumed in the nonlinear gyrokinetic description; while the calculation of collisional fluxes in arbitrary geometry is analyzed in detail in several other works,  \emph{e.g.},  Ref. \citenum{angioni2000neoclassical}. In particular, following the gyrokinetic field theory approach\cite{brizard2007foundations},  below we  introduce the pull-back representation of the particle distribution function in gyrocenter coordinates. This allows us to express the moments of the distribution function, appearing in the fluctuation induced fluxes expression, in terms of the gyrocenter distribution function which, for example, can be computed by means of a gyrokinetic code (cf. Ref. \citenum{garbet2010gyrokinetic} for a recent review). As a result, 
the moment equation approach to transport equations \cite{hinton1976theory} combined with 
the conceptual framework of nonlinear gyrokinetic theory and the push-forward representation of particle moments \cite{frieman1982nonlinear,brizard2007foundations}, allows a compact and physically transparent derivation of cross-field particle and energy fluxes, which include collisional and fluctuation induced transport processes on the same footing \cite{falessi2017gyrokinetic}.

Following Ref. \citenum{brizard2007foundations}, the particle distribution function can be expressed in terms of the guiding-center distribution function $F${\cb{,}} which, in turn, can be written in terms of the gyrocenter distribution function $\bar{F}$:
  \begin{eqnarray}
  \label{eq:102}
  f= e^{- {\bm \rho}\cdot {\bm \nabla}}F  =&  e^{- {\bm \rho}\cdot {\bm \nabla}} \bar F  - \frac{e}{m} e^{- {\bm \rho}\cdot {\bm \nabla}}\left \langle \delta \psi_{gc} \right \rangle \left( \frac{\partial \bar F}{\partial \cal E} + \frac{1}{B_0} \frac{\partial \bar F}{\partial \mu} \right) +
  \left[ \frac{e}{m} \delta \phi \frac{\partial \bar F}{\partial \cal E}  \right] + \\  \nonumber
   &+  \left[  \frac{e}{m} \left( \delta \phi - \frac{v_\parallel}{c} \delta A_\parallel \right)  \frac{1}{B_0} \frac{\partial \bar F}{\partial \mu} + \delta \bm A_\perp \times \frac{\bm b}{B_0} \cdot \bm \nabla \bar F \right]
   \; ,
  \end{eqnarray}
  where  ${\cal E} = v^2/2$ is the energy per unit mass, $\mu$ is the magnetic moment adiabatic invariant $\mu = v_\bot^2/(2 B_0) + \ldots$ and:
  \begin{equation}
  \label{eq:103}
  \delta \psi_{gc} = \delta \phi_{gc} - \frac{\bm v}{c} \cdot \delta \bm A_{gc} = e^{\bm \rho \cdot \bm \nabla} \left( \delta \phi - \frac{\bm v}{c} \cdot \delta \bm A \right) \equiv e^{\bm \rho \cdot \bm \nabla} \delta \psi.
  \end{equation}
  The gyrophase average $\left\langle \delta \psi_{gc}\right\rangle$ involves  introducing Bessel functions as integral operators:
\begin{equation}
\label{eq:104}
\left\langle \delta \psi_{gc}\right\rangle  =  \hat I_0  \left( \delta \phi - \frac{v_\parallel}{c} \delta A_\parallel \right) + \frac{m}{e} \mu \hat I_1  \delta B_\parallel {\cb{\; ,}}
\end{equation}
  where $\hat I_n (x) \equiv (2/x)^n J_n(x)$, $J_n(x)$ are the Bessel functions, $\lambda^2 \equiv 2 (\mu B_0/\Omega^2) k_\perp^2$ and the definition of $\hat I_n$ acting on a generic function $g(\bm r) = \int \hat g (\bm k) \exp (i \bm k \cdot \bm r) d \bm k$ is  the following:
  \begin{equation}
  \label{eq:105}
  \hat I_n g (\bm r) \equiv \int d \bm k e^{i \bm k \cdot \bm r} \hat I_n(\lambda) \hat g (\bm k) .
  \end{equation}
  At the leading order in the asymptotic expansion (with $\delta=\rho/L$ as expansion parameter), we can show that:
  \begin{equation}
  \label{eq:106}
  \left\langle e^{-\bm \rho \cdot \bm \nabla} (...) \right\rangle = \hat I_0 (...) \; ; \;\;\;\; \left\langle e^{-\bm \rho \cdot \bm \nabla} \bm v (...) \right\rangle = \hat I_0  v_\parallel \bm b (...) + \frac{mc}{e} \mu  \hat I_1  \bm b \times \bm \nabla (...).
  \end{equation}
Denoting velocity space integration as $\langle \dots \rangle_{v}$ and using these relations we can show that:
\begin{eqnarray}
\label{eq:107}
\left\langle f \right\rangle_v  & = &
\left \langle \hat I_0 \left[ \bar F - \frac{e}{m} \left(  \frac{\partial \bar F}{\partial \cal E} + \frac{1}{B_0} \frac{\partial \bar F}{\partial \mu} \right) \left\langle \delta \psi_{gc} \right\rangle \right] \right\rangle_v + \\
 \nonumber &  +& \frac{e}{m} \left\langle \frac{\partial \bar F}{\partial \cal E} \right\rangle_v \delta \phi + \frac{e}{m} \left\langle \frac{1}{B_0} \frac{\partial \bar F}{\partial \mu} \left( \delta \phi - \frac{v_\parallel}{c} \delta A_\parallel \right) \right\rangle_v  +  \delta \bm A_\perp \times \frac{\bm b}{B_0} \cdot \bm \nabla \left\langle \bar F \right \rangle_v  \;\;
\end{eqnarray}
\begin{eqnarray}
\label{eq:136a}
\hspace{-7em}\left\langle \bm v_\perp f \right\rangle_v  =  \frac{mc}{e} \bm b \times \left\langle \mu \hat I_1 \bm \nabla \left[  \bar F - \frac{e}{m} \left(  \frac{\partial \bar F}{\partial \cal E} + \frac{1}{B_0} \frac{\partial \bar F}{\partial \mu} \right) \left\langle \delta \psi_{gc} \right\rangle \right]\right\rangle_v \;\;.
\end{eqnarray}
By means of Eq. (\ref{eq:107}), we can compute the leading order of the first term on the right hand side of Eq. (\ref{eq:101}). In particular, we obtain the following 
expression:
\begin{eqnarray}
\label{eq:109}
- \left\langle (en \bm E - en \bm E_0 ) \cdot R^2 \bm \nabla \phi \right\rangle_\psi & = & e \left \langle\left( \frac{1}{c} \frac{\partial}{\partial t} \delta \bm A + \bm \nabla \delta \phi \right) \cdot R^2 \bm \nabla \phi \left\langle f \right\rangle_v \right\rangle_\psi=
\nonumber \\ & = & e \left \langle R^2 \bm \nabla \phi \cdot \bm \nabla \delta \phi   \left\langle \hat I_0 (\lambda) \delta \bar G \right\rangle_v \right\rangle_\psi
\; ,
\end{eqnarray}
where we have introduced the function $\bar G$ \cite{frieman1982nonlinear}:
$$\bar G = \bar F - \frac{e}{m} \frac{\partial \bar F}{\partial \cal E} \left\langle \delta \psi_{gc} \right\rangle {\cb{\; ,}}$$
which satisfies the Frieman-Chen nonlinear gyrokinetic equation up to ${\cal O}(\delta)$\cite{brizard2007foundations,chen2016physics}. 
By means of Eq. (\ref{eq:136a}), and proceeding in the same way with formal manipulations of the second term on the right hand side of Eq. (\ref{eq:101}), after some calculations shown in Appendix \ref{appA}, we obtain:
\begin{eqnarray}
\label{eq:110}
& &\left\langle (en /c) \bm V \cdot (\bm B_0 - \bm B) \times R^2 \bm \nabla \phi \right\rangle_\psi = \left\langle (e/c) \left\langle \bm v f \right\rangle_v \cdot R^2 \bm \nabla \phi \times  (\bm \nabla \times \delta \bm A)  \right\rangle_\psi =\\
\nonumber & & \hspace*{1cm} = e \left\langle\left\langle \frac{v_\parallel}{c} \frac{\bm \nabla \psi \cdot \delta \bm B_\perp}{B_0} \hat I_0 (\lambda) \delta \bar G \right\rangle_v \right\rangle_\psi +
\nonumber\\ & & \hspace*{1.2cm} - m \left\langle\left\langle   \left( \delta B_\parallel R^2 \bm \nabla \phi - \frac{F}{B_0}
\delta \bm B_\perp \right) \cdot  \mu \hat I_1 (\lambda)  \bm \nabla_\perp  \delta \bar G \right\rangle_v \right\rangle_\psi
\nonumber= \\ & & \hspace*{1cm} =
\nonumber e \left\langle R^2 \bm \nabla \phi \cdot \left\langle \bm \nabla  \left( - \frac{v_\parallel}{c} \delta A_\parallel \right) \hat I_0 (\lambda) \delta \bar G
+ \bm \nabla  \left( \frac{m}{e} \mu \delta B_\parallel \right) \hat I_1 (\lambda) \delta \bar G \right\rangle_v \right\rangle_\psi.
\end{eqnarray}
Details of the derivation of Eq. (\ref{eq:110}) are also given in \ref{appA}. Here, it is crucial to stress that, for consistency 
with the transport analysis of Secs. \ref{sec:particletransp} and \ref{sec:energytransp}, based on the moment approach, the contribution to particle
fluxes in Eqs. (\ref{eq:109}) and (\ref{eq:110}) must be understood as ``effectively averaged'' consistently with the spatiotemporal scales of the reference state.

From these expressions, recalling that $\wb{\nabla}\phi \cdot \wb{\nabla} \sim \partial_{\phi}$, we can see that the fluctuation induced transport is due only to toroidally symmetry breaking perturbations, as expected. Meanwhile, the push forward expression for the energy fluxes are identical to the density fluxes except for the weighting factor $m v^{2}/2$, which multiplies every term, as demonstrated in Sec. \ref{sec:energytransp}. Again,
we note that the expressions for fluctuation induced fluxes and ensuing transport are valid for generic short-wavelength turbulence;
that is, for drift wave fluctuations at frequencies much lower than the cyclotron frequency but wavelength as short as the particle 
Larmor radius. Nonetheless, our moment approach is based on an asymptotic expansion, which assumes that the 
effect of fluctuation induced transport is given for structures that are sufficiently longer scale than the Larmor radius. In other words,
although drift-wave turbulence is described by nonlinear gyrokinetic theory, its effect on transport is accounted for on the
length scale typical of the plasma equilibrium. This assumption has been used in the derivation of Eqs. (\ref{eq:101}) and Eq. (\ref{eq:129}) several times, \emph{e.g.}, neglecting terms with the partial time derivative of momentum/energy density.
As already stated, these results generalize and extend the analysis of Ref. \citenum{hinton1976theory}. At the same time, our results recover those originally proposed in Refs. \citenum{plunk2009theory,barnes2009trinity,abel2013multiscale} derived assuming a systematic spatiotemporal scale separation between dynamically evolving plasma equilibrium and turbulent fluctuation spectrum. In fact, by introduction of suitable radial and time averages, Refs. \citenum{plunk2009theory,barnes2009trinity,abel2013multiscale} compute the slow evolution of smoothed equilibrium density and pressure profiles. Our approach{\cb{,}} instead, based on moment equations and nonlinear gyrokinetic theory,  focuses on the modification to the reference state on sufficiently long (spatiotemporal) scales only, without introducing any averaging operation. In other words, for reasonably smooth (or macroscopic) fluctuations, the expressions  we derive hold point-wise in time and space, contrary to the results obtained in Refs. \citenum{plunk2009theory,barnes2009trinity,abel2013multiscale}. It is to be expected that the spatiotemporal average description of Refs.\citenum{plunk2009theory,barnes2009trinity,abel2013multiscale} and our novel approach are consistent. We can verify this by substituting the pull-back representation of the distribution function, Eq. (\ref{eq:102}), into Eq. (A.20) of Ref. \citenum{plunk2009theory}, which describes the transport of particles analogously to Eq. (\ref{eq:116}). The same correspondence can be verified with Eq. (A.25) of Ref. \citenum{plunk2009theory}, which describes energy transport analogously to Eq. (\ref{eq:154}), obtaining, up to the required order, the averaged version of the equations already derived by means of the moment method.

\section{Longer timescales}
\label{sec:longer}
In the previous sections,  we have calculated collisional and fluctuation induced transport processes in an axisymmetric tokamak plasma, using nonlinear gyrokinetic theory. A number of other works dealing with the same problem exist in the literature, \cite{sugama1996transport,balescu1990anomalous,shaing1988neoclassical} and  this issue has been analyzed in depth even more recently \cite{abel2013multiscale}. In all these works, including the present one, collisional and turbulent fluxes are calculated up to $\mathcal{O}(\delta^{2})$ in the asymptotic expansion. Using the characteristic length and time-scales of a modern magnetic fusion device, we can estimate the corresponding time-scale of validity of transport equations, which is of the order of seconds. This is relatively short when compared with the expected duration of a pulse in the next generation Tokamaks, \emph{i.e.} ITER, which is $>3000 s$ \cite{janeschitz2001plasma}. Therefore,
even if a set of equations for the fluxes with a precision of $\mathcal{O}(\delta^{2})$ is not enough in order to predict the behavior of the plasma during a whole ITER plasma discharge, it could be used for the real time control of plasma dynamic evolution based on properly designed actuators (see, \emph{e.g.}, Refs. \citenum{gravesinternal,graves2012control}). Nonetheless,
 it is quite obvious to address collisional and fluctuation induced fluxes up to $\mathcal{O}(\delta^{3})$ 
in order to have predictive simulations of the extremely long time scales of plasmas in fusion reactors. Thus, transport equations are needed with higher accuracy, and the  distribution function, \emph{e.g.}, is needed with an accuracy of $\mathcal{O}(\delta^{2})$. Furthermore, plasma and the surrounding material components need to be considered as ``integrated'' and ``complex'' system with extremely diverse spatiotemporal scales. For example, the characteristic length and time-scales considered in this work typically apply to the core region of thermonuclear plasmas. Generally addressing the problem of the plasma transport as the edge plasma region is approached, where equilibrium magnetic field is modified from closed to open field lines, poses severe issues. In fact, the relative ordering of spatiotemporal scale of turbulent fluctuation spectra and  transport phenomena is also modified and not so well separated as in the plasma core.  In particular{\cb{,}} the radial gradient scale length can be of the same order of the banana width of thermal ions (see, \emph{e.g.}, Ref. \citenum{chang2004numerical}) in the pedestal region, where plasma profile are characterized by sharp variations. Therefore, conventional neoclassical transport theory cannot be applied. For these reasons the study of higher order terms of the asymptotic expansion may be of crucial importance. Thus, studying formal expressions of particle, momentum and energy fluxes that are valid on the time scale of an ITER discharge requires the parallel development of a gyrokinetic theory correct, at least, up to $\mathcal{O}(\delta^{2})$ and of a corresponding more accurate form of collisional fluxes. 

Gyrokinetics is based on an asymptotic theory, where the expansion parameter is defined as the ratio between the gyroradius and the characteristic length scale of variation of the equilibrium magnetic field. This is achieved in two steps: first, the fluctuating electromagnetic fields are ignored and only the background (equilibrium non-uniform) magnetic field  is considered; then, the turbulent fields are introduced and the corresponding plasma responses are calculated. Each step is based on an asymptotic expansion done with different perturbation parameters, which are respectively denoted, see e.g. Ref. \citenum{brizard2007foundations}, by $\epsilon_{B}$ and $\epsilon_{\delta}$. The gyrokinetic ordering typically assumes $\epsilon_{B}\sim\epsilon_{\delta}\sim \delta$. The asymptotic expansion in $\epsilon_{\delta}$ needs to be carried out at least at  up to second order to obtain an energy like invariant. Meanwhile, terms of order $\epsilon_{B}^{2}$ are usually neglected in practical applications because of their complexity  and, therefore, the gyrokinetic ordering is not carried out
on an equal footing with respect to fluctuation intensity and equilibrium magnetic field non-uniformity \cite{brizard2007foundations}. Generally, this is justified as $\epsilon_{B}$ is typically smaller than $\epsilon_{\delta}$ in cases of practical interest. Nonetheless, this issue is known in the fusion research community and efforts are being carried out to derive more accurate perturbation expansions based on gyro-kinetics, which may be applied on longer time scales, \emph{e.g.} Ref. \citenum{tronko2016second}, or in plasma conditions where expansion parameters underlying the asymptotic theory may be not as small as in typical burning plasma core region. This is, e.g., the situation of fusion plasmas in the edge region, as anticipated above, where the presence of material walls surrounding the core plasma volume and of sharp spatial gradients may challenge the standard approach to gyrokinetic theory \cite{xu2007edge}. In general, the perturbative expansions have been consistently carried out up to the second order in $\delta$ only in the electrostatic case, \emph{i.e.} when the turbulent fluctuation spectrum does not significantly affect the magnetic field \cite{parra2011phase}. The more general case of a fully electromagnetic fluctuation spectrum in non-uniform toroidal plasmas has not been addressed to date. Therefore, the  extended form of the pullback of the distribution function, Eq. (\ref{eq:102}), up to the second order in $\delta$ has not been given. This is mainly a technical issue. In fact, it is worthwhile reminding that, in principle, the non-canonical perturbation theory \cite{cary1983noncanonical} allows to formally derive the desired  pullback operator at any order of the asymptotic expansion \cite{brizard2007foundations}. However, the calculation becomes very convoluted already for the second order electrostatic case.
Extending transport theory on long time scales also affects the analysis of Coulomb collisions. 
It is well known \cite{hinton1976theory,balescu1988transport} that neoclassical as well as classical transport theory deal with a linear collision operator which approximates the Landau collision integral. These theories show that the approximated collision operator is consistent with a positive production of entropy and the Onsager symmetry\cite{onsager1931reciprocal} in the linear relations connecting the thermodynamic forces and the fluxes. These are linear closure relations and, therefore, they have a clear interpretation in terms of non-equilibrium thermodynamics. In the study of higher order terms of the asymptotic expansion,  we need to deal, in general, with nonlinear closure relations. In transport theory, the nonlinear closure relations and the (nonlinear) Landau collision operator have been studied with different approaches. In particular, in Ref.\cite{chang2004numerical} this problem has been addressed by means of numerical simulations, while an analytic approach has been carried out by G.Sonnino in a series of works,  see Refs. \citenum{sonnino2007geometrical,sonnino2007minimum,sonnino2008nonlinear,sonnino2009nonlinear,sonnino2014thermodynamic} and the more recent Ref. \citenum{PhysRevE.94.042103}. This author introduces and describes the Thermodynamic Field Theory as a useful tool to derive corrections to the linear closure relations with applications to plasma physics.
\section{Conclusions}
  \label{sec:concl}
In this work we have analyzed particle and energy transport on the energy confinement time scale in a magnetized plasma taking into account the contributions of Coulomb collisions and fluctuations on the same footing. 
  These equations hold at every point in space and do not involve any radial averaging operation. However, our approach assumes that plasma profiles evolve on the macro-scales only; thus, some sort of spatiotemporal averaging is embedded in our approach at this level. This is one difference with the previous works on this topic, \emph{i.e.} Refs. \citenum{abel2013multiscale,plunk2009theory,barnes2009trinity}, based on the systematic scale separation between fluctuating and equilibrium quantities. Another element of novelty is the derivation technique, which uses the moment approach \cite{hazeltine2003plasma} and the gyrokinetic push-forward representation of the fluid moments \cite{brizard2007foundations}. 
  As noted in Sec.\ref{sec:fluctindfluxes}, Eqs. (\ref{eq:109}) and (\ref{eq:110}) as well as the corresponding equation for energy fluxes must be understood as ``effectively averaged'' consistently with the spatiotemporal scales of the reference state. This yields to naturally introducing the notion of spatiotemporal scales of equilibrium variations and of the corresponding structures, which must be self-consistently determined by nonlinear gyrokinetic theory. More precisely, some of the dynamic evolution of the reference state that is produced by the fluctuation spectrum and is not consistent with the well-known transport ordering adopted here \cite{hinton1976theory,frieman1982nonlinear} must be considered separately, as distortion of the reference state itself. Such a distortion is, however, compatible (as it should be) with the gyrokinetic ordering \cite{frieman1982nonlinear} and generally consists of long-lived (undamped by collisionless dissipation processes) phase space structures with meso-spatiotemporal scales. These ``phase space zonal structures'' \cite{zonca2015nonlinear} play a crucial role in transport processes of collisionless fusion plasmas, as they are a measure of the deviation of the system from the local thermodynamic equilibrium\cite{chen2007nonlinear,zonca2015nonlinear,chen2016physics}; that is, from the considered reference state. Furthermore, their dynamic evolution can be secular and eventually invalidate the scale separation assumption between fluctuation and reference state. This issue is discussed in Ref. \citenum{falessi2017gyrokinetic} and will be further analyzed in a future publication. In summary, we note that the moment equation approach illuminates the possibility of providing a unified theory of collisional and fluctuation induced transport by means of a compact and intuitive formulation. It however fails where mesoscale structures become increasingly more important and a kinetic approach is required \cite{falessi2017gyrokinetic}.

\section*{Acknowledgment}
\label{sec:ack}
This work was carried out within the framework of the EUROfusion Consortium and received funding from the Euratom research and training programme 2014-2018 under Grant Agreement No. 633053 (project WP17-ER/ENEA-10).

\appendix

\section{Derivation of the fluctuation induced particle flux}
\label{appA}
The expression for the particle flux induced by fluctuations of the magnetic field derived by means of the moment method, \emph{i.e.} Eq. (\ref{eq:109}), require to evaluate the following expression in terms of the gyrocenter distribution function:
\begin{equation*}
\left\langle (e/c) \left\langle \bm v f \right\rangle_v \cdot R^2 \bm \nabla \phi \times  (\bm \nabla \times \delta \bm A)  \right\rangle_\psi.
\end{equation*}
We can re-write this expression using Eq. (\ref{eq:136a}):
\begin{align*}
  &\sav{\frac{e}{c} \vav{\hat{I}_{0}\left[  \bar{F} - \frac{e}{m} \gav{\delta \psi_{gc}}\frac{\partial \bar{F}}{\partial \mathcal{E}}\right]v_{\parallel} \wb{b}}\cdot R^{2}\wb{\nabla}\phi\times \left(\wb{\nabla}\times\delta \wb{A}  \right)} +\\
+  & \sav{\vav{ m \mu \hat{I}_{1} \wb{b}\times \wb{\nabla} \delta \bar{G}}\cdot R^{2} \wb{\nabla}\times
(\delta B_{\parallel}\wb{b} + \delta \wb{B}_{\perp})}.
\end{align*}
Using the following identity:
\begin{equation*}
\wb{b}\cdot \wb{\nabla} \phi\times \delta\wb{B}= B_{0}^{-1}R^{-2}\wb{\nabla}\psi\cdot \delta \wb{B}
\end{equation*}
we can re-write the first term as:
\begin{equation*}
\sav{\vav{\frac{e v_{\parallel}}{c}\hat{I}_{0}\delta \bar{G}}\frac{\wb{\nabla}\psi\cdot \delta\wb{B}}{B_{0}}}.
\end{equation*}
At the leading order $\delta B_{\perp}=\wb{\nabla}\delta A_{\parallel} \times \wb{b}$ and, therefore, we can write:
\begin{equation*}
\sav{\vav{\frac{e v_{\parallel}}{c}\hat{I}_{0}\delta \bar{G}} B_{0}^{-1}\wb{\nabla}\psi\cdot(\wb{\nabla}\delta A_{\parallel}\times \wb{B}) }.
\end{equation*}
Using the identity $\wb{b}\times \wb{\nabla}\psi= F \wb{b} - R^{2}B \wb{\nabla}\phi$ we finally obtain:
\begin{equation*}
-e \sav{\vav{\wb{\nabla}\left( \frac{\delta A_{\parallel}v_{\parallel}}{c} \right)\hat{I}_{0} \delta \bar{G}}\cdot R^{2} \wb{\nabla}\phi}.
\end{equation*}
The second term to calculate is the following:
\begin{equation}
\label{eq:appdc3a}
\sav{\vav{ m \mu \hat{I}_{1} \wb{b}\times \wb{\nabla} \delta \bar{G}}\cdot R^{2} \wb{\nabla}\times
(\delta B_{\parallel}\wb{b} + \delta \wb{B}_{\perp})}
\end{equation}
which is the sum of two contributions. We can show that:
\begin{equation*}
  \wb{b}\times \wb{\nabla}\delta \bar{G}\cdot R^{2} \wb{\nabla}\phi \times \delta \wb{B}_{\perp}= \frac{F}{B_{0}}\delta \wb{B}_{\perp}- \wb{\nabla}_{\perp}\delta \bar{G}
\end{equation*}
and, therefore, we can re-write the second term of Eq. (\ref{eq:appdc3a}):
\begin{equation}
\label{eq:appdc3}
\sav{\vav{m \mu \frac{F}{B_{0}} \delta \wb{B}_{\perp}\cdot \hat{I}_{1}\wb{\nabla}_{\perp}\delta \bar{G}}}.
\end{equation}
Analogously we can show that $(\wb{b}\times \wb{\nabla} \delta\bar{G})\cdot (\wb{\nabla}\phi \times \wb{b}) = - \wb{\nabla}\phi \cdot \wb{\nabla}_{\perp}\delta \bar{G}$ and we can re-write the first term of Eq. (\ref{eq:appdc3a}) as:
\begin{equation}
\label{eq:appdc5}
-m \sav{\vav{\left(\delta B_{\parallel}R^{2}\wb{\nabla}\phi- \frac{F}{B_{0}}\delta\wb{B}_{\perp}\right)\cdot \mu \hat{I}_{1}\wb{\nabla}_{\perp}\delta \bar{G}}}.
\end{equation}
This is again the sum of two terms. The first one:
\begin{equation*}
-m \sav{\vav{\delta B_{\parallel}R^{2}\wb{\nabla}\phi \cdot \mu \hat{I}_{1}\wb{\nabla}_{\perp}\delta \bar{G}}}
\end{equation*}
can be written at the leading order as:
\begin{equation*}
-m \sav{\vav{\delta B_{\parallel}R  \mu \hat{I}_{1} \frac{\partial \delta \bar{G}}{\delta \phi}}}.
\end{equation*}
This can be written, noting that the surface average involves an average over the angular coordinate $\phi$, as:
\begin{equation}
m \sav{ \vav{ R^{2} \wb{\nabla}\phi \cdot \wb{\nabla}(\delta B_{\parallel}\mu) \hat{I}_{1}\delta \bar{G}}}.
\end{equation}
The second term of Eq. (\ref{eq:appdc5}) can be rewritten as:
\begin{equation*}
  m\sav{\vav{\frac{F}{B_{0}}\frac{\wb{\nabla}\psi}{R^{2}B_{0}^{2}}\cdot \left(\wb{\nabla} \delta A_{\parallel}\frac{\partial}{\partial \phi}\delta \bar{G} - \wb{\nabla} \delta \bar{G}\frac{\partial}{\partial \phi} \delta A_{\parallel} \right) }}  
\end{equation*}
which can be neglected with respect to others, being of higher order in a large aspect-ratio tokamak.

\bibliography{updatedbib.bib}
\end{document}